\documentclass[11pt,english]{revtex4-1}
\usepackage[T1]{fontenc}
\usepackage[latin9]{inputenc}
\usepackage{amstext}
\usepackage{graphicx}

\makeatletter
\@ifundefined{textcolor}{}
{%
 \definecolor{BLACK}{gray}{0}
 \definecolor{WHITE}{gray}{1}
 \definecolor{RED}{rgb}{1,0,0}
 \definecolor{GREEN}{rgb}{0,1,0}
 \definecolor{BLUE}{rgb}{0,0,1}
 \definecolor{CYAN}{cmyk}{1,0,0,0}
 \definecolor{MAGENTA}{cmyk}{0,1,0,0}
 \definecolor{YELLOW}{cmyk}{0,0,1,0}
}

\makeatother

\usepackage{babel}

\begin{document}

\title{Proton beams with controlled divergence and concentrated energy in
TNSA regime by USUI laser pulse interaction with a tailored hole-target}

\author{Huan Wang$\,^1\,^2\,^3$, Lihua Cao$\,^1\,^3\,^4$, and X. T. He$\,^1\,^3\,^4\,^5$}

\affiliation{$^1$Center for Applied Physics and Technology, Peking University, Beijing
100871, China }

\affiliation{$^2$Institute of Plasma Physics and Fusion, Peking University, Beijing
100871, China }

\affiliation{$^3$Key Laboratory of High Energy Density Physics Simulation (HEDPS)
of the Ministry of Education, Peking University, Beijing 100871, China }

\affiliation{$^4$Institute of Applied Physics and Computational Mathematics, Beijing,
100088, China }

\affiliation{$^5$Institute for Fusion Theory and Simulation, Zhejiang University,
Hangzhou, 310027, China }

\begin{abstract}
An improved acceleration scheme to produce protons with controlled
divergence and concentrated energy density is studied using ultrashort ultraintense (USUI) laser pulse
interaction with a tailored hole-target in target normal sheath acceleration (TNSA) regime. Two-dimension-in-space and three-dimension-in-velocity (2D3V) particle-in-cell (PIC) simulations show that the tailored hole-target helps to reshape the sheath electric field and generate a transverse quasistatic electric field of $TV/m$ along the inner wall of the hole. The transverse electric field suppresses the transverse expansion
of the proton beam effectively, as it tends to force the produced protons
to focus inwards to the central axis, resulting in controlled divergence and concentrated
energy density compared with that of a single plain target. The dependence of proton beam divergence and energy feature on depth of the hole is investigated in detail. A rough estimation of the hole depth ranges depending on $a_{0}$ of the incident laser is also given.
\end{abstract}
%\email{lihua_cao@iapcm.ac.cn}
\email{wang.huan@pku.edu.cn}
%\email{xthe@iapcm.ac.cn}

\maketitle

\section{INTRODUCTION}

The laser-driven ion acceleration from ultrashort ultraintense (USUI)
laser pulse interaction with various solid targets has been studied actively
for applications ranging from basic particle physics {[}1{]},
bench-top particle accelerators{[}2{]}-{[}4{]}, medical therapies{[}3{]},
fast ignition of inertial controlled fusion (ICF){[}4{]}{[}5{]}, etc. Up
to now, several mechanisms for accelerating ions have been proposed,
such as target normal sheath acceleration(TNSA){[}2{]}{[}8{]}, laser
breakout after-burner(BOA){[}9{]}, and radiation pressure acceleration(PRA){[}10{]}{[}11{]}.
Many potential applications require proton and ion beams with high
collimation, monoenergetic, larger particle number and intense
energy density, as a result, the enhancement of beam quality becomes
of intriguing interest and numerous experimental and theoretical
studies have been devoted to achieve this goal{[}3{]}{[}12{]}-{[}17{]}.

In TNSA regime, which is more stable than other ions acceleration mechanisms, when an
intense laser pulse irradiates on a thin plain target, a large number
of electrons are accelerated by the incident laser and then transport
to the target backside, forming an electron cloud and thus a strong
electrostatic charge-separation field there. A population of protons
near the rear surface of the target then are pulled out and accelerated
into the backside vacuum by this so-called target normal sheath electric field{[}2{]}{[}18{]}.
One of the challenges here is how to generate a nearly local and uniform
sheath field at the target rear surface, as typically the sheath field
occupies a large area and is inhomogeneous due to the divergent electrons which establish the sheath field, and more to the point, when
the normally incident laser is p-polarized, in which case the electric field of the laser oscillates along the transverse direction and thus pushes electrons to move up or down to either lateral edges of the plain target and assemble there, as a consequence, the target normal sheath electric field is stronger on both the lateral edges symmetrically while relatively much weaker on the central axis, and this edge effect will get greater over time. In order to control
the shape of the sheath electric field and then to obtain a collimated
proton beam with high quality, several tailored structural targets and
ions-doped foil targets are proposed previously{[}19{]}{[}20{]}.

In this paper, we study a practical scheme to generate proton beams
with controlled divergence and concentrated energy in TNSA regime by
a USUI laser illuminating the tailored hole-target, and we also
employ a same single plain target without the backside hole as
a comparison. Two-dimension-in-space and three-dimension-in-velocity (2D3V) particle-in-cell (PIC) simulations demonstrate
the effectiveness of the tailored hole-target in suppressing the transverse
proton beam divergence by confining the sheath electric filed in
the hole almost locally and uniformly and also by generating a transverse electric field
to focus the protons. Accordingly, the dependence of the proton beam
characteristics on the target hole depth is also worthy to
investigate in detail.

This work is organized as follows. In Sec. II, we present the target
model and the simulation parameters. For comparison, both of the single
plain and tailored hole-target are considered. In Sec. III,
our simulation results are presented, from which one can see the robust
improvement on the beam divergence and protons energy intensity by using the tailored hole-target. The dependence of proton beam divergence on depth of the hole is examined thoroughly in Sec. IV, as well as some estimations about hole depth range on laser intensity $a_{0}$. A conclusion is given in Sec. V.

\section{\label{{Sec:Sec2}}Target configuration and simulation parameters}

\begin{figure}[h!]
\begin{centering}
\includegraphics[scale=0.5]{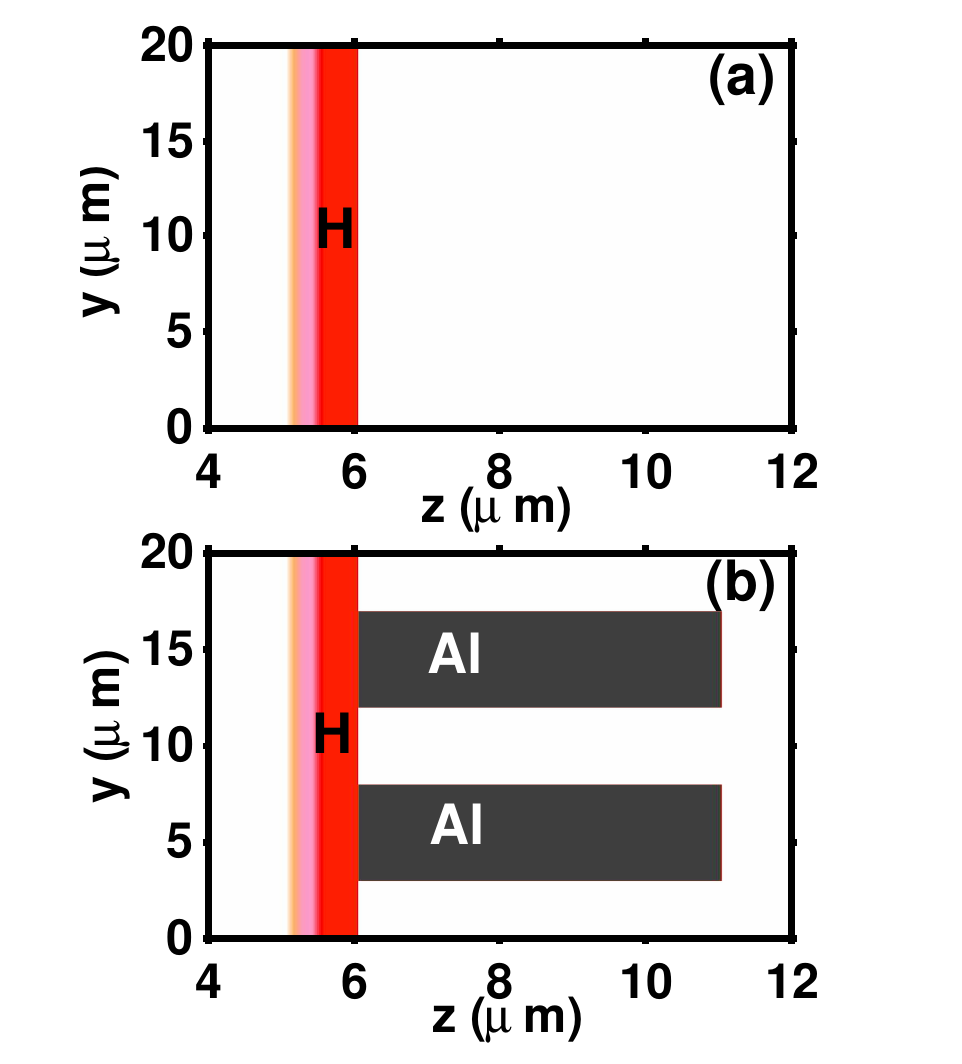}
\par\end{centering}

\protect\caption{\label{figure:Fig1}{\footnotesize{}The conceptual diagram of the single plain target
and the tailored hole-target studied in our simulation. (a) the single
plain target with the initial electrons and protons peak density of $50n_{c}$,
and (b) the tailored hole-target, which is designed to have the same
configutation and particle settings with the single plain one, and
together with these, a hole with depth of $h=5\mu m$ and diameter
of $4\mu m$ surrounding by two $5\mu m$ thick, $100n_{c}$ dense
horizontal ramparts made up of $Al^{3+}$ and electrons. The front
side of the plain target is located at $z=5\mu m$, and its thickness
and width is $1\mu m$ and $20\mu m$, respectively. In order to
include the prepulse effect, we employ a linear density gradient in
$0.5\mu m$ at the laser illumination surface.The initial temperature
of electrons is set to be $1000eV$. The target materials have been marked.}}

\end{figure}

The simulations are performed with a 2D3V PIC code KLAP2D{[}17{]}.
In the simulations, 1500 cells longitudinally along $z-$axis and 2000
cells transversely along $y-$axis constitute a $15\mu m\times20\mu m$
simulation box. A $p-$polarized laser pulse with $a_{0}=6$ (intensity
$I_{0}\approx4.9\times10^{19}W/cm^{2}$) and wavelength $\lambda_{0}=1\mu m$
is normally incident on a solid plain target comprised of electrons
and protons. The front side of the plain target is located at $z=5\mu m$,
of which thickness is $1\mu m$ and width is $20\mu m$. In order
to include the prepulse effect, we employ a linear density gradient
in $0.5\mu m$ at the laser illumination surface. Figure \ref{figure:Fig1}
shows a conceptual diagram of the single plain target and the tailored
hole-target studied in our simulation. In the case of the single plain
target, see Figs. 1(a), the initial electrons and protons peak density
is $50n_{c}$, and the tailored hole-target is designed to have the
same configuration and particle settings with the single plain one,
and together with these, a hole with depth of $h$ and diameter of
$4\mu m$ surrounding by two $5\mu m$ thick, $100n_{c}$ dense horizontal
ramparts made up of $Al^{3+}$ and electrons, see Figs. 1(b).
The initial temperature of electrons is set to be $1000eV$. About
$(1.4\sim2.0\times10^{6})$ superparticles are employed in our simulations.
The laser pulse coming from the left boundary has a transverse Gaussian
profile with beam waist $r_{0}=2.5\mu m$ and a trapezoidal temporal
profile of duration $\tau=22T$, consisting of a plateau of $20T$
and rising and falling periods of $1T$ each, where $T$ is the laser
period. As has been studied by T. P. Yu et al.\cite{key-13} and M.
Nakamura et al.\cite{key-19}, the optimal hole diameter is of the
order of the laser spot size and proton beam divergence and energy characteristics
have little dependence on pulse duration, so we fix the laser
parameters and hole diameter all through this paper.

\section{Effect of hole-target in reshaping sheath field and inducing transverse electric field}

\begin{figure}[h!]
\begin{centering}
\includegraphics[scale=0.6]{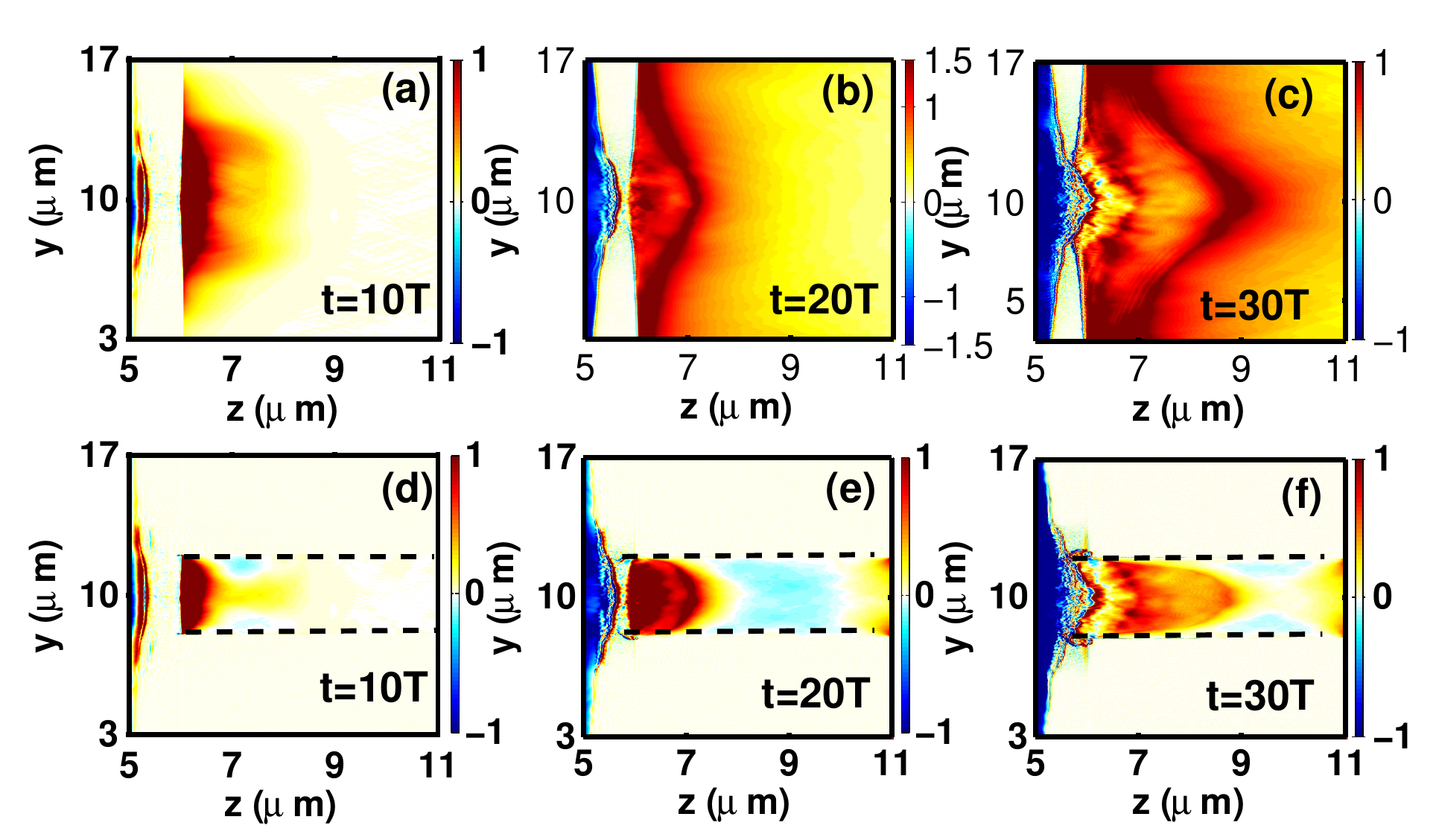}
\par\end{centering}

\protect\caption{\label{figure:Fig2}{\footnotesize{}Distributions of cycle-averaged sheath
electric fields $E_{sheath}$ at $t=10T,\,20T,\,30T$ for plain (a)-(c)
and hole-target (d)-(f), respectively. The dashed black lines show
the initial inner boundaries of the Al ramparts. The electric fields are
normalized by $\frac{m_{e}\omega_{0}c}{e}$, where $m_{e}$ , $\omega_{0}$
, $c$ and $e$ are electron rest mass, laser angular
frequency, light speed in vacuum and electron charge, respectively.}}
\end{figure}

The main acceleration mechanism we consider here is TNSA, which
means the shape of sheath electric field determined by the accelerated
hot electron cloud spread is vital to the quaity of the subsequently
generated proton beam. Firstly, the shape of sheath electric fileds
for both plain and tailored hole-target are investigated. Figs. \ref{figure:Fig2}(a)-(f)
show distributions of cycle-averaged sheath field $E_{sheath}$ ($E_{sheath}$ is actually the longitudinal electric field $E_{z}$)
at $t=10T,\,20T,\,30T$ for plain (top) and hole-target (bottom),
respectively. One can notice that in Figs. \ref{figure:Fig2}(a)-(c)
for the plain target, at early time, say $t=10T$, the maximum sheath
field $E_{sheath}^{max}\sim1.2\times10^{13}V/m$ is centralized around
the laser incident axis with a diameter of about $10.3\lambda$
and the further away from the central axis, the weaker $E_{sheath}$. However,
as the accelerated hot electrons are further exploding into the vacuum, say at $t=20T$,
$E_{sheath}$ is expanding along both the transverse and longitudinal
directions as well, resulting in a wider and longer bell shape. What's
more, $E_{sheath}$ is still y-axial symmetric but is stronger ($E_{sheath}^{edges}\sim6.6\times10^{12}V/m$)
at the target top and bottom edges than that ($E_{sheath}^{axis}\sim4.5\times10^{12}V/m$)
in the region nearer around laser axis. At later time points $T\geq30T$, $E_{sheath}$
has expanded more decentralized, and the maximum sheath field is obviously
located at the top and bottom edges of the plain target, leaving the centraxonial
$E_{sheath}$ much more weaker. In addition, because the incident laser is
p-polarized, the electric field of the laser pushes electrons along
the transverse direction, and thus makes the hot electron denser at the edges.
Add it all up, the transverse edge effect of $E_{sheath}$ is
one of the reasons which lead to proton beam divergence and large spot size. As shown in Figs. \ref{figure:Fig2}(d)-(f), in
the case of the hole-target, the transverse edge effect of $E_{sheath}$
is suppressed by the horizontal ramparts made of Al, which forces $E_{sheath}$
to distribute uniformly and locally in the hole, and from our simulation
results, we obtain $E_{sheath}\sim9.0\times10^{12}V/m,\,4.8\times10^{12}V/m,\,2.1\times10^{12}V/m$
at $t=10T,\,20T,\,30T$, respectively. The reason why the hole can
help to eliminate the edge effect is the accelerated hot electrons can not transport freely in the closely-behind Al ramparts, in which denser electrons
prohibit them from going forward and thus $E_{sheath}$ can not be established in these regions like that of the plain target.

\begin{figure}[h!]
\begin{centering}
\includegraphics[scale=0.5]{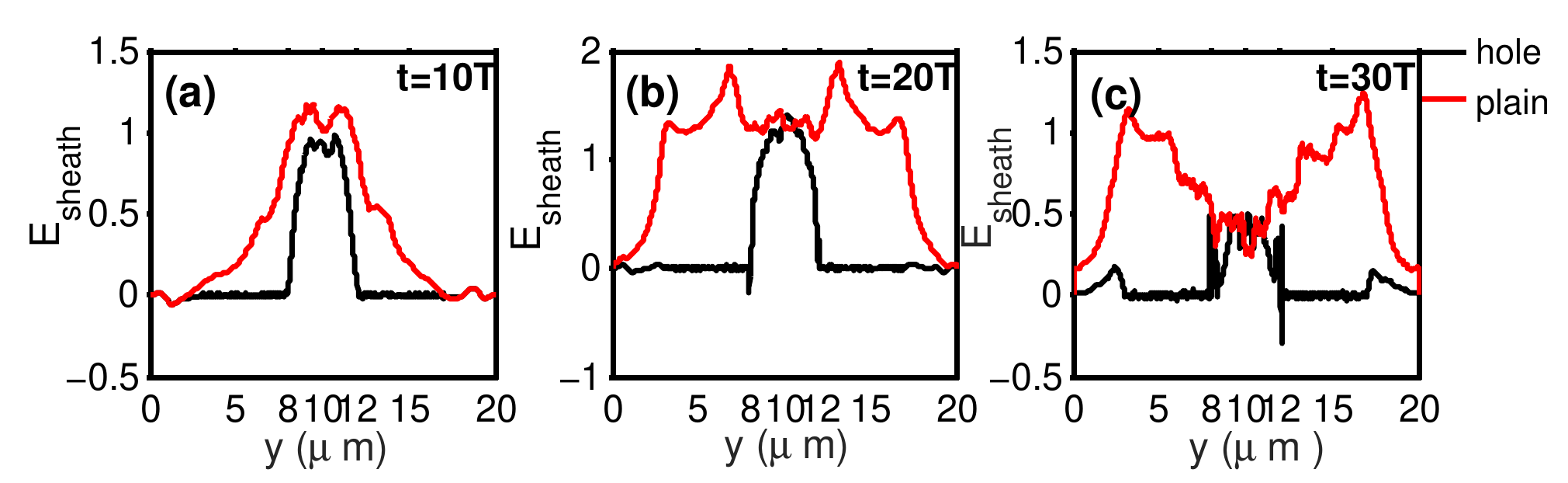}
\par\end{centering}

\protect\caption{\label{figure:Fig3}{\footnotesize{}Sectional view for the crosssection
located at $z=6.5\mu m$ at time points (a) $t=10T$, (b) $t=20T$,
(c) $t=30T$, red line is for the plain target and black line is for
the tailored hole-target. The electric fields $E_{sheath}$ are cycle-averaged
and normalized by $\frac{m_{e}\omega_{0}c}{e}$.}}
\end{figure}

As the shape of $E_{sheath}$ evolves over time, we pick up three
snap shots of sectional view (the crosssection is at $z=6.5\mu m$
) at the same time points mentioned above, and the results are presented
in Figure \ref{figure:Fig3}. Compared with $E_{sheath}$ of the plain
target (red line), which is much broader and has two sharp corners
almost symmetrical about the central axis ($\text{y=10\ensuremath{\mu}m}$),
$E_{sheath}$ of the tailored hole-target (black line) is limited just
tightly around central axis, being local and uniform. Moreover, for
the single plain target, as we discussed above, the two corners of
$E_{sheath}$ are moving outwards to the top and bottom edges of the
plain target and meantime the central part of $E_{sheath}$ is getting weaker.
The maximum $E_{sheath}$ shown in Figure \ref{figure:Fig3} is
smaller for the tailored hole-target due
to less accelerated hot electrons produced in the hole than that in vacuum of the plain target rear side.

\begin{figure}[h!]
\begin{centering}
\includegraphics[scale=0.6]{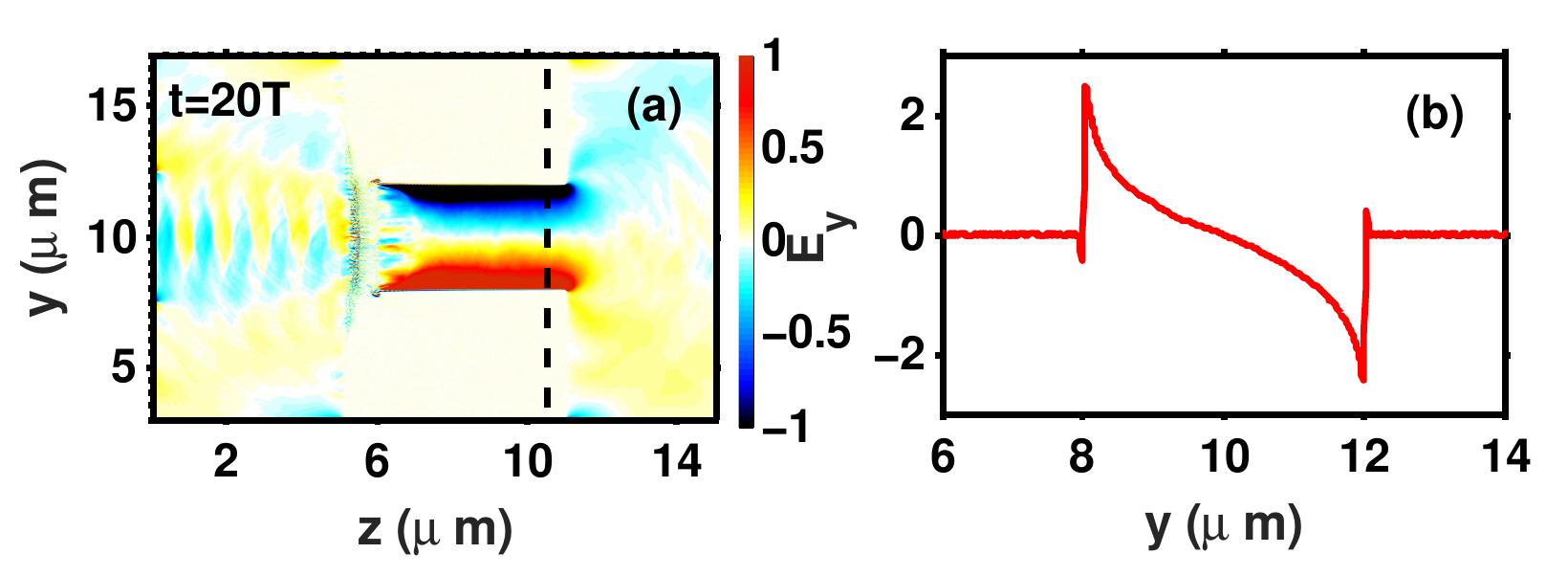}
\par\end{centering}

\protect\caption{\label{figure:Fig4}{\footnotesize{}(a) Distribution of transverse electric
field $E_{y}$ in the hole at $t=20T$; (b) sectional view of transverse electric field $E_{y}$ for the crosssection
located at $z=11\mu m$ at $t=20T$, where the black dashed line indicates the crosssection. The electric fields
$E_{y}$ are cycle-averaged and normalized by $\frac{m_{e}\omega_{0}c}{e}$.}}
\end{figure}

The tailored hole-target not only has an advantage over the plain
one in controlling the shape of $E_{sheath}$, but also has a transverse
electric field $E_{y}$ induced in the hole which does not exist in the case
of the plain target. As one can see in Figure \ref{figure:Fig4},
$E_{y}$ has opposite directions in the hole, that is, in the upper
side $E_{y}$ is negatively along $-y$ while in the lower side $E_{y}$
is positively along $+y$, thus this transverse field tends to focus and confine the protons in the transverse direction as a tight bunch. The transverse electric field can reach as large as $6.43\times10^{12}V/m$ according to our simulation results, which is approximately consistent with $$E_{y}\sim\frac{\Phi_{y}}{e\lambda_{D}}\sim\frac{(\sqrt{1+a_{0}^{2}}-1)m_{e}c^{2}}{e\lambda_{D}}=6.71\times10^{12}V/m$$
where we estimate $\lambda_{D}=\sqrt{\epsilon_{0}T_{e}/n_{e}e^{2}}$ as the Debye length of accelerated hot electrons with temperature $T_{e}\approx m_{e}c^{2}\sqrt{1+\frac{2 U_{p}}{m_{e}c^{2}}}\sim 2.23MeV$.

\begin{figure}[h!]
\begin{centering}
\includegraphics[scale=0.5]{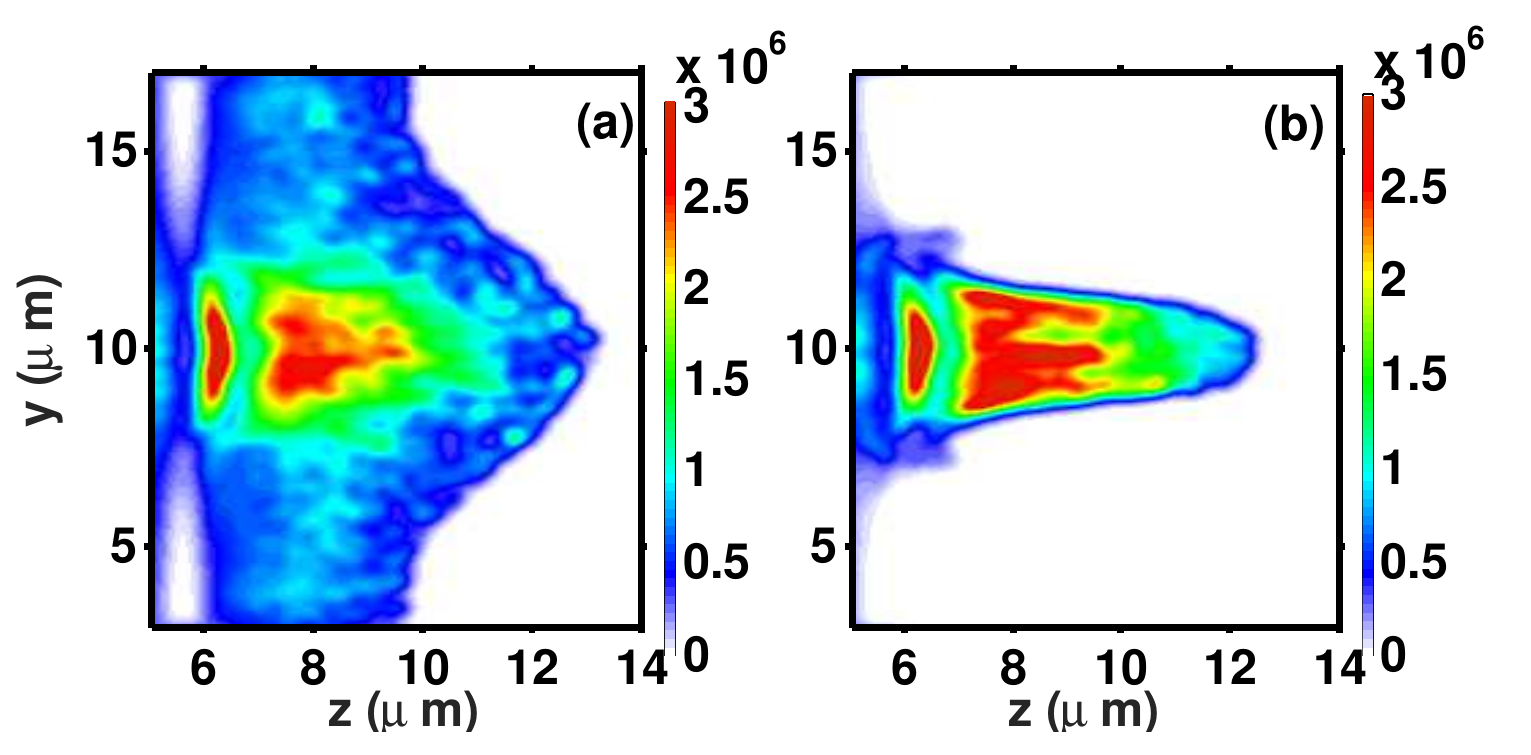}
\par\end{centering}
\protect\caption{\label{figure:Fig5}{\footnotesize{}Distributions of proton energy density in the region $z\geq5\mu m$ at $t=30T$ for (a) the plain target, (b) tailored hole-target. Protons energy is in unit of $m_{e}c^{2}$.}}
\end{figure}

In order to have a clear understanding about the energy distribution characteristics of generated protons in the rear side of the targets, we use Figure \ref{figure:Fig5} to show the distribution of proton energy density in the region beyond $z=5\mu m$ at time $t=30T$ for both plain and tailored hole-target. One can ensure that thanks to both the reshaped $E_{sheath}$ and the induced transverse electric field $E_{y}$, protons produced by the tailored hole-target have been confined in the hole with energy more greatly concentrated compared with that of the plain target.

\section{the dependence of proton beam divergence on depth of the hole and depth range estimation}

\begin{figure}[h!]
\begin{centering}
\includegraphics[scale=0.5]{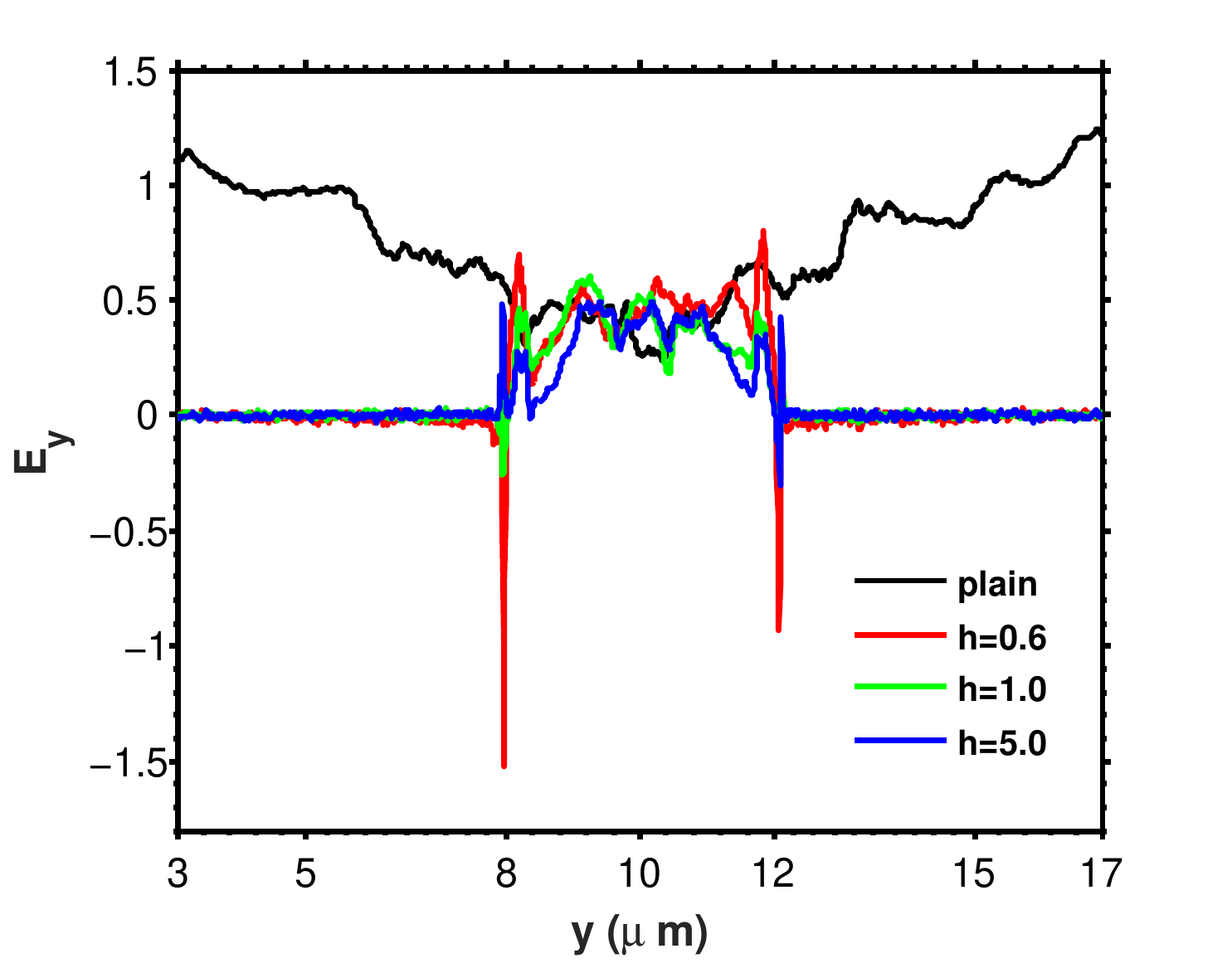}
\par\end{centering}
\protect\caption{\label{figure:Fig6}{\footnotesize{}Distribution of rear sheath electric fields $E_{sheath}$ for the single plain target and tailored hole-target with different depth $h=0.6\mu m, 1\mu m, 5\mu m$ at $t=30T$, where the black line is for the plain target, red, green and blue lines are for $h=0.6$, $h=1$ and $h=5$, respectively. The electric fields $E_{sheath}$ are cycle-averaged and normalized by $\frac{m_{e}\omega_{0}c}{e}$.}}
\end{figure}

In Section II when declaring the simulation parameters, we fix diameter of the hole $(D=4\mu m)$ and leave depth $h$ as the only geometric variable for good reason. Here we investigate the dependence of proton beam divergence and energy feature on $h$. Figure \ref{figure:Fig6} demonstrates the rear sheath field $E_{sheath}$ at $t=30T$ for the single plain target and tailored hole-target with different depth $h=0.6\mu m, 1\mu m, 5\mu m$, which are given respectively by lines in different colors as stated in legend. As we have proved above, there is severe edge effect for the plain target, leading to proton divergence. All the hole-targets have their sheath electric fields limited in the inner hole, except for the difference that the field distributions are distinct marginally (at hole inner margin $z=8\mu m$ and $z=12\mu m$) and in the hole. In the case of $h=0.6\mu m$, marginal sheath field is the strongest among the four targets, which may result in more accelerated electrons in marginal region and thus large beam divergence; and further more, "marginal effect" makes $E_{sheath}$ fluctuate more wildly in the hole. For $h=1\mu m$ case, we still see the "marginal effect", but a little eased. The most optimal shape of sheath electric field for accelerating protons is in the case of $h=5\mu m$, where the marginal fields are almost suppressed to equal the level of the electric field in the hole, and therefore, the sheath field $E_{sheath}$ is reshaped to be uniform and centralized. From our simulation results, we ensure that $h=5\mu m$ is the optimal depth parameter among these cases, for too shallow the "marginal effect" affects $E_{sheath}$ significantly.

\begin{figure}[h!]
\begin{centering}
\includegraphics[scale=0.5]{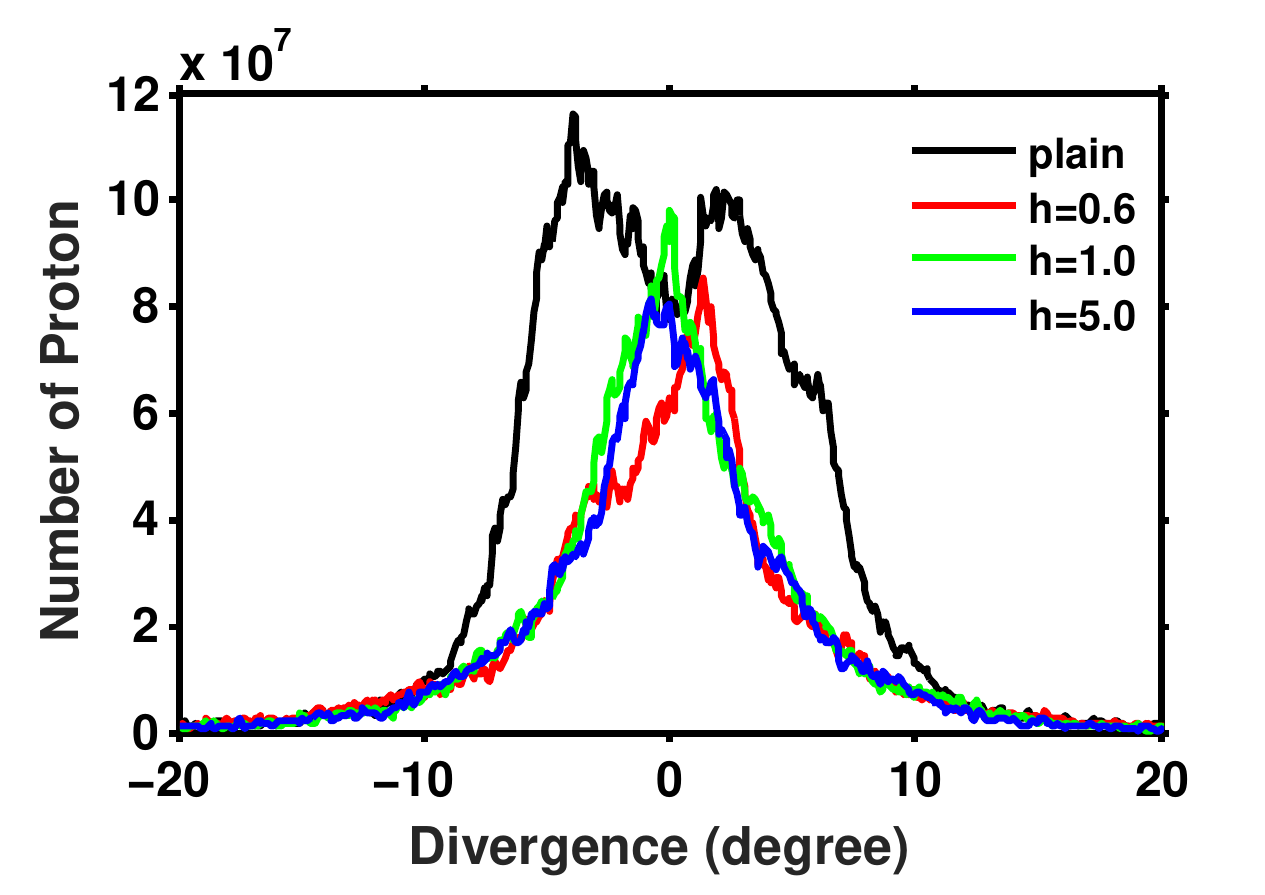}
\par\end{centering}
\protect\caption{\label{figure:Fig8}{\footnotesize{}Divergence spectrum of accelerated forward-going protons in the rear side at $t=30T$ for the single plain target and tailored hole-target with different depth, where the black line is for the plain target, red, green and blue lines are for $h=0.6\mu m$, $h=1\mu m$, and $h=5\mu m$, respectively. The divergence angle is in unit of degree, and the proton number is indicated in the vertical axis.}}
\end{figure}

Now we focus on the dependence of proton beam divergence on depth of the hole. Figure \ref{figure:Fig8} shows the divergence spectrum of accelerated forward-going protons in the rear side at $t=30T$ for the single plain target and tailored hole-target with different depth $h=0.6\mu m, 1\mu m, 5\mu m$. Here divergence angle is defined by the following formula $$\theta_{div}=\arctan{\frac{p_{y}}{p_{z}}}$$
where $p_{y}$ and $p_{z}$ are protons transverse and longitudinal relativistic momentum respectively. One can see clearly that the proton beam from the plain target has two divergence angle peaks in $\theta_{div}=3.946^{\circ}$ and $\theta_{div}=1.91^{\circ}$, while proton beam from the hole-target with $h=0.6\mu m$ has an obvious angular deviation from the central axis and one angle peak in $\theta_{div}\approx1.32^{\circ}$, and proton beams from the hole-targets with $h=1\mu m$ and $h=5\mu m$ each has angle peaks in $\theta_{div}\approx0.04^{\circ}$ and $\theta_{div}\approx0.19^{\circ}$ respectively. Although the hole-target with $h=1\mu m$ has a smaller peak divergence angle, on the whole its divergence spectrum shape is fatter than hole-target with $h=5\mu m$. The accelerated proton number in the case $h=5\mu m$ from our simulation results is about $9.65\times10^{9}$.

\begin{figure}[h!]
\begin{centering}
\includegraphics[scale=0.5]{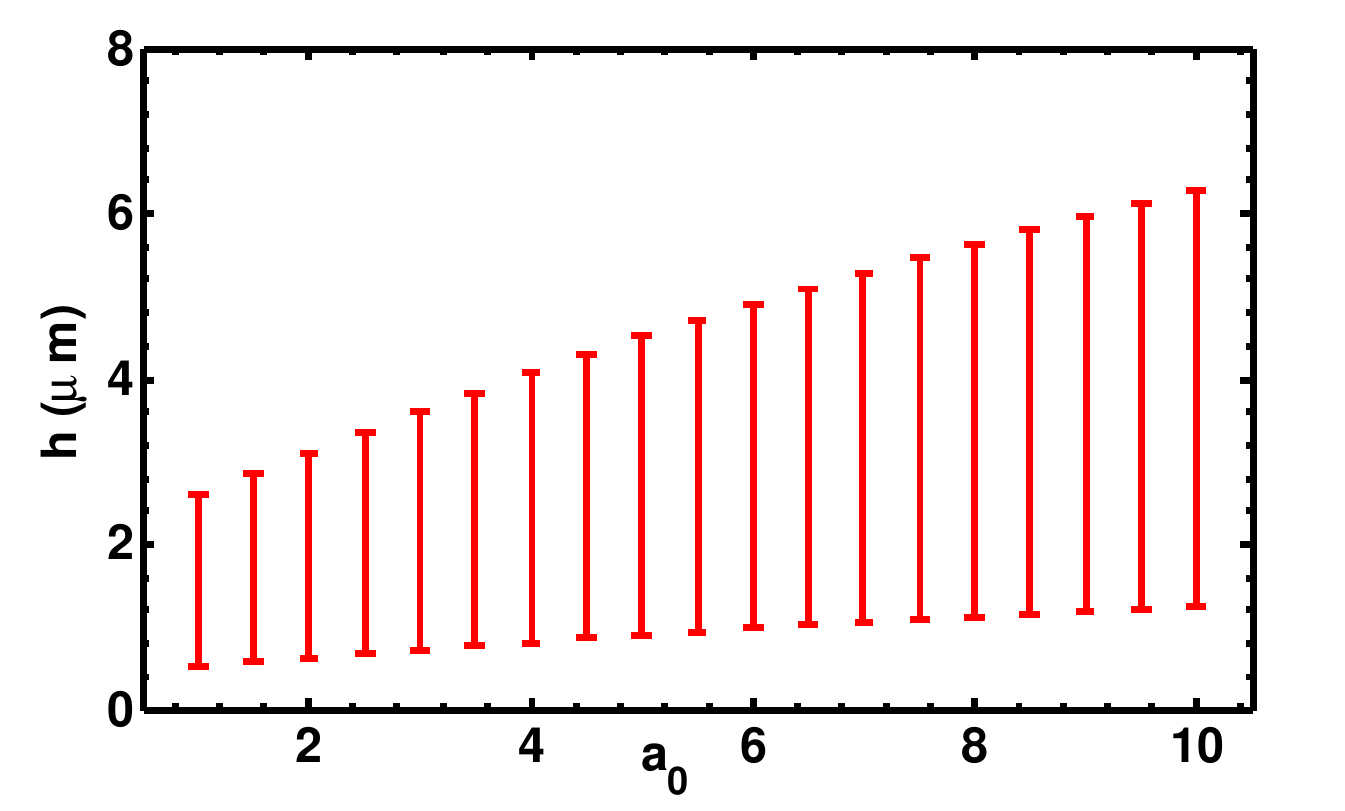}
\par\end{centering}
\protect\caption{\label{figure:Fig9}{\footnotesize{}A rough estimation of the hole depth ranges depending on $a_{0}$ of the incident laser.}}
\end{figure}

So far, we have learned the primary function of the ramparts of the hole-target is to eliminate the electric sheath field "edge effect", confine the accelerated hot electrons and focus the subsequent protons, and we have demonstrated the dependence of proton beam divergence on depth of the hole, and next, the depth range should be figured out. As we know, the accelerated hot electrons temperature is $T_{e}\approx m_{e} c^{2}(1+\frac{1}{2}a_{0}^{2})^{1/2}$, and according to the following formula $$T_{e}= eE_{l}L_{n}$$ we can estimate the minimum depth of the hole should be the local plasma scale length $L_{n}=C_{s}t\sim 0.47(1+\frac{1}{2}a_{0}^{2})^{1/4}$, where $C_{s}=\sqrt{T_e/m_{i}}$ is the ion sound speed, $E_{l}$ is the longitudinal accelerating field and $t$ is taken to be the incident laser duration. Now we make some assumptions about the maximum value of $h$. Wilks et al. found that protons gain energy proportional to the electron temperature{[}2{]}, $$E_{p}=\alpha T_{e}$$
where $\alpha$ is somewhere between 2 and 12, depending on the model, and from our simulation results, we can roughly estimate $\alpha\approx5$. As a result, we can obtain $h_{max}\sim \alpha C_{s}t\approx 2.35(1+\frac{1}{2}a_{0}^{2})^{1/4}$. Figure \ref{figure:Fig9} shows the rough range of hole depth, and we can see using our simulation parameters, $h_{min}\approx0.98\mu m$ and $h_{max}\approx4.91\mu m$. It is reasonable that for incident laser with large $a_{0}$, one should use hole-targets with longer ramparts, and the minimum depth shows little change.

\section{Conclusion}

In conclusion, proton acceleration in TNSA regime using a tailored hole-target has been investigated by 2D3V PIC simulations. It is found that proton beams with more intense energy density and much smaller divergence angle can be produced from suitably picked hole depth compared with those from the single plain target. The dependence of proton beam divergence on depth of the hole is also investigated. We give a rough estimation of the depth ranges depending on $a_{0}$ of the incident laser, and all the simulation results demonstrate that as long as the depth of the hole roughly satisfies the effective action range of the hole target ramparts, the tailored hole-target works effectively.

\end{document}